\documentclass[twocolumn, secnumarabic,amssymb, aps]{revtex4-1}

\usepackage[pagewise]{lineno}

\usepackage[utf8]{inputenc}
\usepackage{CJKutf8}
\usepackage{amsmath,amssymb}
\usepackage{type1cm}
\usepackage{mathrsfs}
\usepackage{mathtools}
\usepackage{url}
\usepackage{graphicx}
\usepackage{color}
\usepackage{booktabs}
\usepackage{braket}
\newcommand\HL[1]{{\color{black}#1}}

\newcommand\HLPOF[1]{{\color{black}#1}}

\newcommand\HLJPCO[1]{{\color{black}#1}}
\usepackage{bm}
\newcommand{\vect}[1]{\boldsymbol{\mathbf{#1}}}
\def\vec#1{\vect{#1}}

\begin{document}
\title{Three-dimensional analysis of vortex-lattice formation in rotating Bose--Einstein condensates using smoothed-particle hydrodynamics}
\author{Satori Tsuzuki}
\email[Email:~]{tsuzukisatori@g.ecc.u-tokyo.ac.jp}
\author{Eri Itoh}
\email[Email:~]{eriitoh@g.ecc.u-tokyo.ac.jp}
\author{Katsuhiro Nishinari}
\email[Email:~]{tknishi@mail.ecc.u-tokyo.ac.jp}

\affiliation{Research Center for Advanced Science and Technology, University of Tokyo, 4-6-1, Komaba, Meguro-ku, Tokyo 153-8904, Japan}
\affiliation{Research Center for Advanced Science and Technology, University of Tokyo, 4-6-1, Komaba, Meguro-ku, Tokyo 153-8904, Japan}
\affiliation{Department of Aeronautics and Astronautics, School of Engineering, The University of Tokyo, 7-3-1, Hongo, Bunkyo-ku, Tokyo, 113-8656, Japan}

\begin{abstract}
Recently, we presented a new numerical scheme for vortex lattice formation in a rotating Bose--Einstein condensate (BEC) using smoothed particle hydrodynamics (SPH) with an explicit time-integrating scheme; our SPH scheme could reproduce the vortex lattices and their formation processes in rotating quasi-two-dimensional (2D) BECs trapped in a 2D harmonic potential. In this study, we have successfully demonstrated a simulation of rotating 3D BECs trapped in a 3D harmonic potential forming ``cigar-shaped'' condensates. We have found that our scheme can reproduce the following typical behaviors of rotating 3D BECs observed in the literature: (i) the characteristic shape of the lattice formed in the cross-section at the origin and its formation process, (ii) the stable existence of vortex lines along the vertical axis after reaching the steady state, (iii) a ``cookie-cutter'' shape, with a similar lattice shape observed wherever we cut the condensate in a certain range in the vertical direction, (iv) the bending of vortex lines when approaching the inner edges of the condensate, and (v) the formation of vortex lattices by vortices entering from outside the condensate. Therefore, we further validated our scheme by simulating rotating 3D BECs.
\end{abstract}
\maketitle

\section{Introduction}
Vortex lattice formation in the Bose--Einstein condensation (BEC) is one of the most interesting phenomena in condensed matter physics. They have been observed in many related problems in low-temperature physics, for example, quantum fluids such as liquid helium-4 and ultracold dilute gases. Recently, we proposed a Lagrangian numerical scheme~\cite{10.1063/5.0143556} to reproduce vortex lattices in rotating BECs using an explicit time-integrating scheme and smoothed particle hydrodynamics (SPH)~\cite{gingold1977smoothed, doi:10.1146/annurev.aa.30.090192.002551}, which was initially proposed in astrophysics for n-body problems such as interactions among galaxies and has been widely used to solve weakly compressible flows. \HL{To our knowledge}, we are the first to employ SPH to spatially discretize the Gross--Pitaevskii (GP) equation \HL{for rotating BECs}.

Because the GP equation is a nonlinear Schr${\rm \ddot{o}}$dinger equation of interacting bosons, its Hamiltonian is described as a many-body interaction system~\cite{Rogel-Salazar_2013, Salasnich2017BSIUA, doi:10.1063/5.0122247}. Therefore, compared to other Euler methods such as finite element (FE)~\cite{VERGEZ2016144, doi:10.1137/15M1009172, HEID2021110165} and finite difference (FD)~\cite{PhysRevE.62.1382, PhysRevE.62.2937, PhysRevE.62.7438} methods, the mechanical picture of the SPH scheme is closer to real BEC physics; we can describe the GP equations as a many-body interaction system in SPH. Therefore, we expect that the microscopic interactions among atomic particles can be replicated more accurately using analytical particles in the simulation domain. \HL{In addition, our recent study showed that the SPH simulation of the angular momentum-conserving two-fluid model incorporating vortex dynamics reproduces vortex lattices~\cite{doi:10.1063/5.0060605}. Our another study provided a \HLPOF{theory} that the equation of motion for inviscid fluids in the two-fluid model becomes equal to the \HLPOF{quantum} hydrodynamic equation derived from the GP equation under specific conditions in SPH form~\cite{doi:10.1063/5.0122247}. Therefore, describing the GP equation in SPH form may help to find some connection with these studies.}
In Ref.~\cite{10.1063/5.0143556}, we successfully reproduced typical cases of vortex lattices in rotating quasi-two-dimensional (2D) BECs by discretizing the GP equation using SPH. We found that our SPH scheme can produce the shapes of vortex lattices and their formation processes observed in previous simulations~\cite{PhysRevA.67.033610} and experiments~\cite{PhysRevLett.86.4443} of rotating BECs by other groups with a certain level of accuracy. 

Our previous study focused on a simulation to reproduce rotating quasi-2D BECs trapped in a 2D harmonic potential to form disc-shaped condensates. \HL{However}, the proposed model is theoretically applicable to rotating 3D BECs because we derived a generic form of the SPH discretization of the GP equation in our previous work~\cite{10.1063/5.0143556}. In this study, we have demonstrated our scheme by simulating rotating 3D BECs trapped in a 3D harmonic potential and forming the ``cigar-shaped'' condensates. Although the numerical scheme was completed in our previous work, realizing a 3D simulation enables us to further validate our SPH scheme. Furthermore, we can explore the 3D structure of vortex lattices and the effect\HL{s} of spatial condensate oscillations and spatial vortex interactions on vortex lattice formation. To this end, the first application of SPH to rotating 3D BECs is scientifically significant and novel and is a strong motivation for writing this paper.

The remainder of this paper is organized as follows. Section 2 briefly reviews SPH and the scheme presented in Ref.~\cite{10.1063/5.0143556}. We introduce an improved boundary treatment that is more suitable for 3D problems. Section 3 presents our simulation results for rotating 3D BECs in a 3D harmonic potential. Section 4 summarizes the paper. 

\section{Methods} \label{sec:methods}
Here, we briefly review our SPH scheme with rephrasing, as necessary for 3D cases. 
The GP equation for a condensate trapped in a 3D harmonic potential and rotated in the horizontal direction is as follows~\cite{PhysRevA.71.063616}:
\begin{eqnarray}
\scalebox{0.85}{$
(i-\gamma)\frac{\partial \psi}{\partial t} = \Biggl[ -\nabla^2 + V + C|\psi|^2 -\mu + i\Omega({x}{\partial_{y}}-{y}{\partial_{x}})\Biggr]\psi,
$}\label{eq:3dGPeq}
\end{eqnarray}
where $\psi$ represents the wavefunction, $\mu$ is the chemical potential, $\gamma$ is the phenomenological dissipation parameter, and $C$ is the coupling constant. The parameter $\Omega$ is a scalar multiple of the angular frequency in the horizontal (x or y) direction, $\omega_{\perp}$. $V$ is the harmonic potential, expressed as $V=[ (1+\epsilon)x^2 + (1-\epsilon)y^2 ]/4 + \lambda^{-1} z^2/4$ in 3D cases, where $\epsilon$ is the anisotropy parameter. $\lambda$ is the ratio of $\omega_{\perp}$ to the angular frequency in the $ z $-vertical direction, $\omega_{z}$. The wavefunction, length, and time in Eq.~(\ref{eq:3dGPeq}) are scaled to $\bar{\psi} = \sqrt{N}a_{h}^{-1} \psi$, $\bar{x} = a_{h}x$, and $\bar{t} = \omega_{\perp}^{-1} t$, respectively. Here, $\bar{x}$, $\bar{t}$, and $\bar{\psi}$ are the variables in the original GP equation. Regarding the operators, $\nabla = (\partial_x, \partial_y, \partial_z)$ and $\partial_{a} \psi$ represent the $a$-component of $\nabla \psi$. Additionally, $\nabla^2 \psi = \{\frac{\partial^2}{\partial x^2}+\frac{\partial^2}{\partial y^2}+\frac{\partial^2}{\partial z^2}\}\psi$. 

In SPH, a discrete physical quantity at a point is described as a continuous quantity by expressing it as a convolution of the Dirac delta function, $\delta$, over the entire domain ($ \small \varphi(\vec{r}) = \int \varphi(\vec{r})\delta(\vec{r} - \vec{\acute{r}}) d\vec{\acute{r}}$), which is approximated by a distribution $W$ that converges to $\delta$ as the effective radius of $W$, $h$, approaches zero; therefore,
\begin{eqnarray}
\scalebox{0.95}{$
\varphi(\vec{r}) \simeq \int \varphi(\vec{r})W(\vec{r} - \vec{\acute{r}}, h) d\vec{\acute{r}}
$}\label{eq:approxbykernel}
\end{eqnarray}
establishes when $\small \lim_{h \to 0} W(\vec{r} - \vec{\acute{r}}, h) = \delta (\vec{r} - \vec{\acute{r}})$. The distribution $W$ and its radius $h$ are referred to as the kernel function'' and ``kernel radius,'' respectively. The integral in Eq.~(\ref{eq:approxbykernel}) is then replaced with a summation as $\small \varphi(\vec{r}_i) = \sum_{j=1}^{N_p} \varphi(\vec{r}_j) {\Delta{V}_j}W_{ij}$, where ${\small \Delta{V}_j}$ is the volume fragment and $N_p$ is the number of discrete points within the effective radius of point $i$. 

The advantage of expressing physical quantities in the SPH form is that they are expressed in a many-particle interaction form. Specifically, a physical quantity at a point in space is obtained as the average of those defined on the other discretization points in the neighbors of the point. We refer to these points as ``particles;'' therefore, a physical quantity $\varphi$ for a particle is expressed as the sum of the contributions from neighboring particles, as expressed in the discrete expression of Eq.~(\ref{eq:approxbykernel}). The gradient and Laplacian of $\varphi$ are derived using the acting operator $\nabla$ in Eqs.~(\ref{eq:approxbykernel}) or its discrete form and then applying the Gauss divergence theorem to the equation systems. We found that the gradient and Laplacian of $\varphi$ were also expressed in many-particle interaction forms. Specifically, the gradient of $\varphi$ is given by \cite{doi:10.1146/annurev.aa.30.090192.002551, 10.1063/5.0143556} 
\begin{eqnarray}
\scalebox{0.95}{$
\nabla \varphi(\vec{r}_{i}) = \sum^{N_{p}}_{j} \Delta{V}_{j} \Bigl[ \varphi(\vec{r}_{i}) + \varphi(\vec{r}_{j}) \Bigr] \nabla W_{ij},
$}\label{eq:SPHgrad}
\end{eqnarray} 
where we use the mathematical relationship $\small \nabla f = \rho[\nabla(f/\rho)+(f/\rho^2)\nabla \rho]$ to represent $\varphi$ in symmetric form~\cite{doi:10.1146/annurev.aa.30.090192.002551} to ensure the action and reaction relationship. A form of the Laplacian of $\varphi$ can be expressed as follows: 
\begin{eqnarray}
\scalebox{0.95}{$
\nabla^{2} \varphi(\vec{r}_{i}) = \sum^{N_{p}}_{j} \Delta{V}_{j} \frac{\varphi(\vec{r}_{i})-\varphi(\vec{r}_{j})}{|\vec{r}_{i}-\vec{r}_{j}|^{2}} (\vec{r}_{i} - \vec{r}_{j})\cdot \nabla W_{ij}.
$}\label{eq:SPHlaps}
\end{eqnarray} 
For more details on SPH, see Sec.~I\hspace{-1.2pt}I.B in Ref.~\cite{10.1063/5.0143556}.
Notably, all $\varphi$, $\nabla \varphi$, and $\nabla^2 \varphi$ are represented in the form of many-particle interactions in SPH.

In the simulations, we divided the wavefunction $\psi$ into real and imaginary parts. By substituting $\psi = u+iw$ into Eq.~(\ref{eq:3dGPeq}), we obtained a pair of real-time dependent equations, one each for $u$ and $w$, corresponding to Eq.~(2) in Ref.~\cite{10.1063/5.0143556}. We discretized each of these time-dependent equations, wherein the gradients $\nabla u$, $\nabla w$ and Laplacians $\nabla^2 u$ and $\nabla^2 w$ were computed using Eqs.~(\ref{eq:SPHgrad}) and (\ref{eq:SPHlaps}) by referring to the particles in the effective interaction range of the kernel function. In this manner, we used SPH only for the discretization of the operators; all particles were fixed at the same locations as in the initial state during the simulations. In addition, we employed the 4th-order Runge--Kutta method, which is an explicit time-integrating scheme. As a treatment of the domain boundaries, in regions that can be regarded as sufficiently far from the condensate, we used a sigmoid function to decay the rotational forces according to the distance from the center, enabling them to be smoothly connected to the domain boundaries where they are zero. 

In quantum mechanics, the square of a wave function at a point indicates the probability that a particle exists at that point~\cite{Berman2018}. Therefore, the integration of the wave function $\psi$ over the entire domain becomes one. Thus, we normalized $u$ and $w$, which were obtained after solving the time-evolution equation to satisfy $\int |\psi|^2 dv=1$. Please refer to Sec.~I\hspace{-1.2pt}I.A and B in Ref.~\cite{10.1063/5.0143556} for further details on the computational method.

\begin{figure*}[t]
\vspace{-35.0cm}
 \includegraphics[width=4.6\textwidth, clip, bb= 0 0 4374 2390]{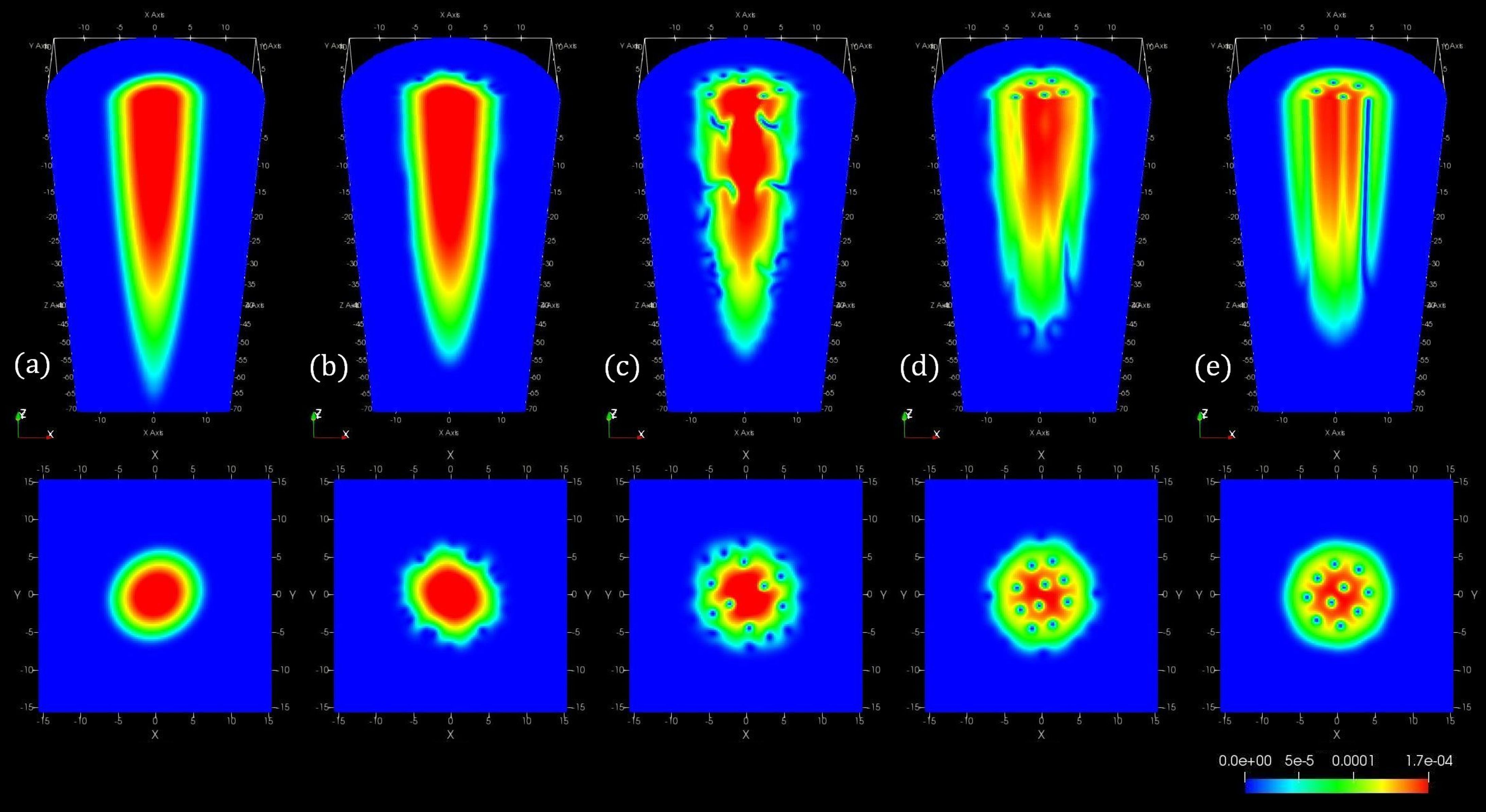}
\caption{Evolution of the density profile on the cross-section at z = 0 from shortly after the start of the rotation to a steady state (lower row) and those on the cross-section of the half cylinder at y = 0 in the region of z $\le$ 0 (upper row).
\HLJPCO{Each pair of snapshots in (a)-(e): the cross-section of the rectangular simulation domain at z = 0 (lower row) and the bird's-eye view of a quarter section of the domain where y = 0 and z $\ge$ 0 (upper row) represent the density profiles at approximately (a) 22, (b) 91, (c) 129, (d) 195, and (e) 367 ms, respectively. We set the parameters $(\gamma, C, N, \lambda, \epsilon, a, a_{h}, \omega_{\perp})$ to $(0.03, 500, 3\times{10}^{5}, 9.2, 0.025, 5.29, 0.728, 2\pi \times 108.56)$, respectively. We set the simulation domain $(L_x, L_y, L_z)$ to (29, 29, 145) to reproduce the geometry of a cigar-shaped potential in Ref.~\cite{PhysRevA.71.063616}. We also set the number of computational particles in each direction $(N_x, N_y, N_z)$ to (116, 116, 580) in proportion to the length of the simulation domain to maintain the same unit length 0.25 in each direction. Because we impose the condition that the density at a constant distance from the center is kept at zero, the region where the density is nonzero becomes cylindrical.}}

\label{fig:Figure1}
\end{figure*}
\begin{figure}[t]
\vspace{-1.6cm}
 \includegraphics[width=0.65\textwidth, clip, bb= 0 0 960 720]{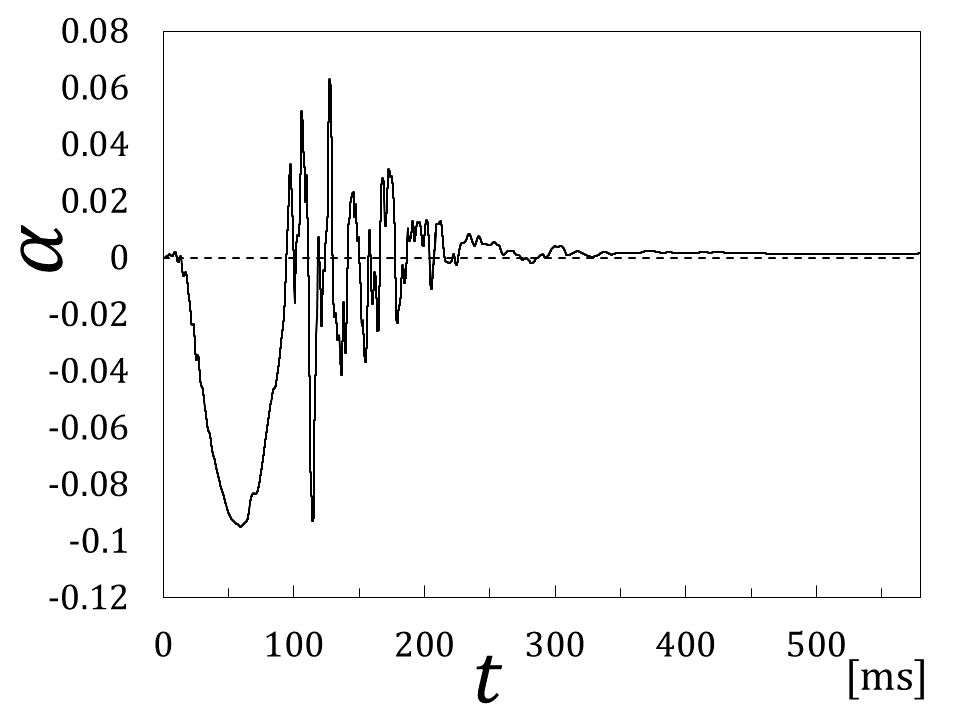}
 \caption{\HLJPCO{Time variation of the deformation parameter $\alpha$ in the case of Fig.~\ref{fig:Figure1}}.}
\label{fig:Figure2}
\end{figure}
\begin{figure*}[t]
\vspace{+0.0cm}
 \includegraphics[width=1.0\textwidth, clip, bb= 0 0 1404 1131]{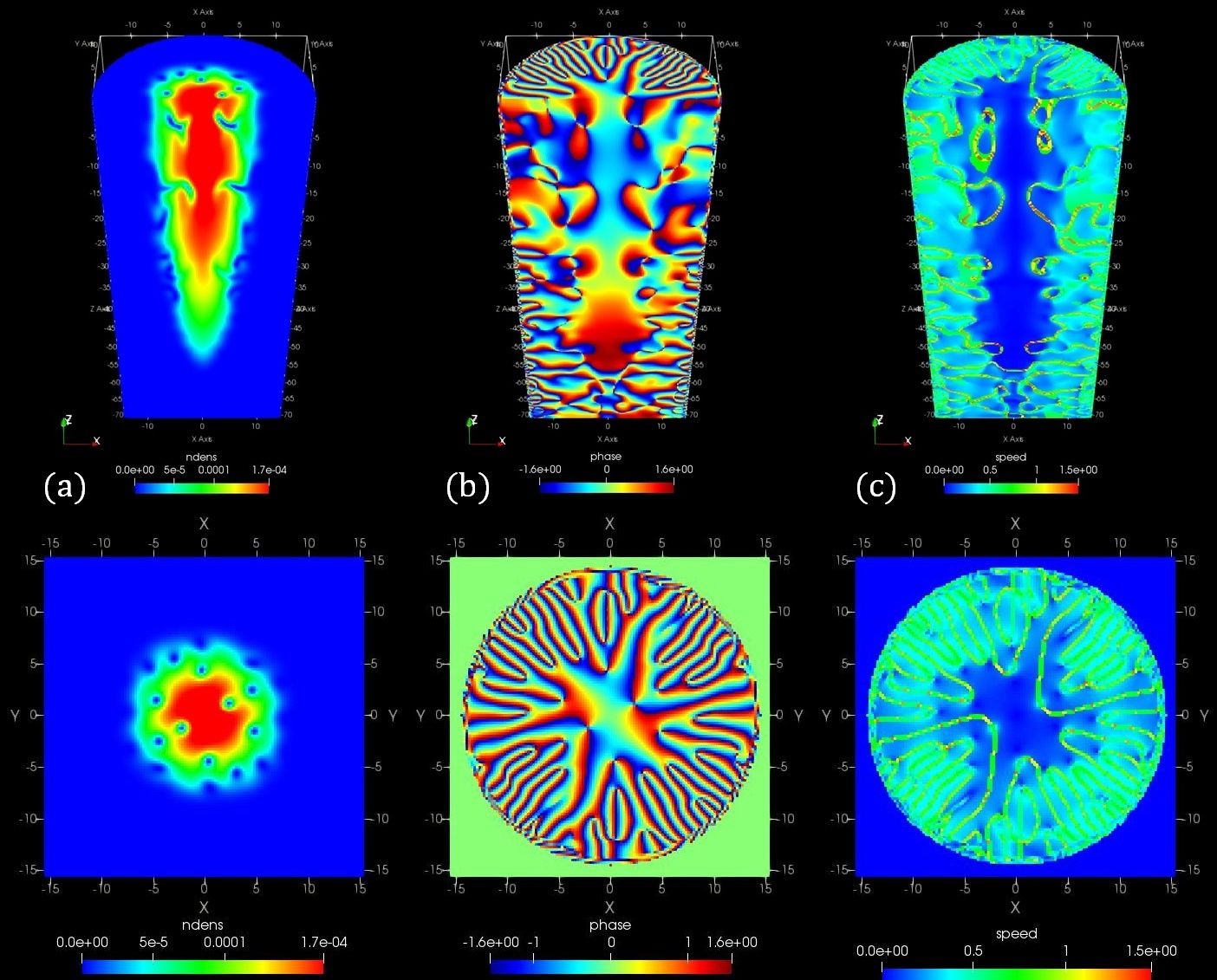}
\caption{Snapshots of the density, phase, and velocity distributions at the time of Fig.~\ref{fig:Figure1}(c).}
\label{fig:Figure3}
\end{figure*}

\section{Numerical analysis}
We used our SPH model to simulate rotating BECs trapped in a 3D harmonic potential, resulting in cigar-shaped condensates. We set the physical parameters following the numerical tests conducted by Kasamatsu et al. using the FD method with an implicit time-integrating scheme reported in Ref.~\cite{PhysRevA.71.063616}. Specifically, we set the parameters $(\gamma, C, N, \lambda, \epsilon, a, a_{h}, \omega_{\perp})$ to $(0.03, 500, 3\times{10}^{5}, 9.2, 0.025, 5.29, 0.728, 2\pi \times 108.56)$. Here, $N$ represents the total number of atomic particles (not the SPH particles). The units of the nondimensionless parameters $a$, $a_{h}$, and $\omega_{\perp}$ are $\rm nm$, $\rm \mu m$, and $\rm Hz$, respectively. 
\HLJPCO{We set the simulation domain $(L_x, L_y, L_z)$ to $(29, 29, 145)$, where $L_z$ was set five times larger than $L_x$ and $L_y$ to reproduce a cigar-shaped potential in a 3D domain reported in the literature~\cite{PhysRevA.71.063616}. We discretized the simulation domain by a unit length of 0.25 in each direction and placed one computational particle (analytical particle) at each discretized point. Hence, the number of analytical particles in each direction, $(N_x, N_y, N_z)$, was $(116, 116, 580)$. These particles remained at the same positions because in this study we solved only the GP equation and there was no advection term as in the fluid equation. Accordingly, $N_{x} \times N_{y} \times N_{z}$ $(116 \times 116 \times 580 = 7804480)$ fixed particles existed within the simulation domain throughout the simulation. Concurrently, we placed boundary particles along the simulation domain whose density was kept zero to ensure that the density distribution was smoothly connected to the area where the density was zero. Finally, this simulation comprised 9041088 particles including boundary particles. Although the number of particles was large, the computational cost was relatively low because we did not have to solve the advection of particles as in the fluid motion simulation.}
All the simulations were performed on a GPU NVIDIA GeForce RTX2080 Ti. We implemented a neighbor-particle list to reduce the computational cost from $\mathcal{O}(N_{a}^2)$ to $\mathcal{O}(N_{a})$ and introduced a linked-list technique to reduce memory usage of the neighbor-particle list~\cite{GREST1989269, GOMEZGESTEIRA2012289}. We set $\delta t$ to $2.5\times 10^{-4}$ and computed approximately 500 ms of the physical time. 

Consequently, we found that our SPH scheme can reproduce the following typical behaviors of rotating 3D BECs reported in the literature~\cite{PhysRevA.71.063616, PhysRevLett.86.4443, 10.1063/5.0143556}: (i) the characteristic shape of the lattice formed in the cross-section at the origin and its formation process, (ii) the stable existence of vortex lines in the vertical direction after reaching the steady state, (iii) a ``cookie-cutter'' shape, where a similar vortex lattice is observed wherever the condensate is cut horizontally in a certain vertical region, (iv) the bending of the vortex line when approaching the inner edges of the condensate, and (v) the formation of vortex lattices by vortices entering from outside the condensate. 

With respect to (i), Fig.~\ref{fig:Figure1} shows the evolution of the density profile on the cross-section at z = 0 from shortly after the start of the rotation to a steady state (lower row) and those on the cross-section of the half cylinder at y = 0 in the region of z $\le$ 0 (upper row). 
\HLJPCO{The upper row in Fig.~\ref{fig:Figure1} shows a bird's eye view from an angle, and thus, the z-axis also appears slanted.}
We confirmed that two key features, the number of vortices after reaching a steady state and the formation process of the vortex lattice, agree with previous simulations~\cite{PhysRevA.71.063616, 10.1063/5.0143556} and experiments~\cite{PhysRevLett.86.4443}. Specifically, as shown in the lower part of Fig.~\ref{fig:Figure1}, the condensates on the cross-section at z = 0 initially oscillate periodically and maintain a stable rotation in a slanted elliptical shape, after which its surface become unstable and ripples are generated, which finally become vortices and penetrate the interior of the condensate, forming a vortex lattice. This is also confirmed from the upper part of Fig.~\ref{fig:Figure1}(c) that vortices enter from outside the condensate. Moreover, the vortices maintain a stable state in a straight form in the vertical direction, as shown in Fig.~\ref{fig:Figure1}(e), which shows that a hollow stretches straight from the vortex of the cross-section at z = 0.


\HLJPCO{Figure~\ref{fig:Figure2} shows the time variation of the deformation parameter, $\alpha$, on the cross-section at z = 0, where $\alpha$ is defined as $\alpha \coloneqq -\Omega \frac{\braket{x^2} - \braket{y^2}}{\braket{x^2} + \braket{y^2}}$. $\braket{A}$ represents the expected value obtained using the formula $\braket{A} = \int A |\varphi|^2 dx dy $ over the cross-section. The results confirm that the vortex lattice reaches a steady state after approximately 350 ms. The shape of the vortex lattice shown in Fig.~\ref{fig:Figure1}(e) and its formation processes (a)--(d) are qualitatively consistent with the numerical experimental results reported in the literature~\cite{PhysRevA.71.063616}}.

Snapshots of the density, phase, and velocity distributions at the time of Fig.~\ref{fig:Figure1}(c) are shown in Fig.~\ref{fig:Figure3}. Video of the entire evolution process for Fig.~\ref{fig:Figure3} is provided as supplementary material, available as an auxiliary file with the preprint version. From the lower rows of Figs.~\ref{fig:Figure3}(a) and (b), we can confirm that the branch point corresponds to a vortex, as observed in the 2D disc-shaped case~\cite{10.1063/5.0143556}. However, as indicated in the upper row of Figs.~\ref{fig:Figure3}(a) and (b), the spatial structures inside the domain from the origin (z = 0) to the bottom of the cylinder change dynamically during the turbulent state, as shown in Fig.~\ref{fig:Figure1}(c), which gradually stabilizes and forms straight hollows in the vertical direction. In this manner, we observed (ii) and (iii), i.e., the stable existence of quantum vortices and the ``cookie-cutter'' shape in the vertical direction after reaching a steady state. The interaction between vortices can be observed in Fig.~\ref{fig:Figure3}(c), which presents the speed distributions calculated from the quantum-mechanical relationship $|\vec{v}| = |\nabla \theta|$ where $\theta$ is the phase given by $\theta = {\rm atan}(w/u)$. 

\begin{figure*}[t]
\vspace{+0.5cm}
 \includegraphics[width=1.0\textwidth, clip, bb= 0 0 1750 923]{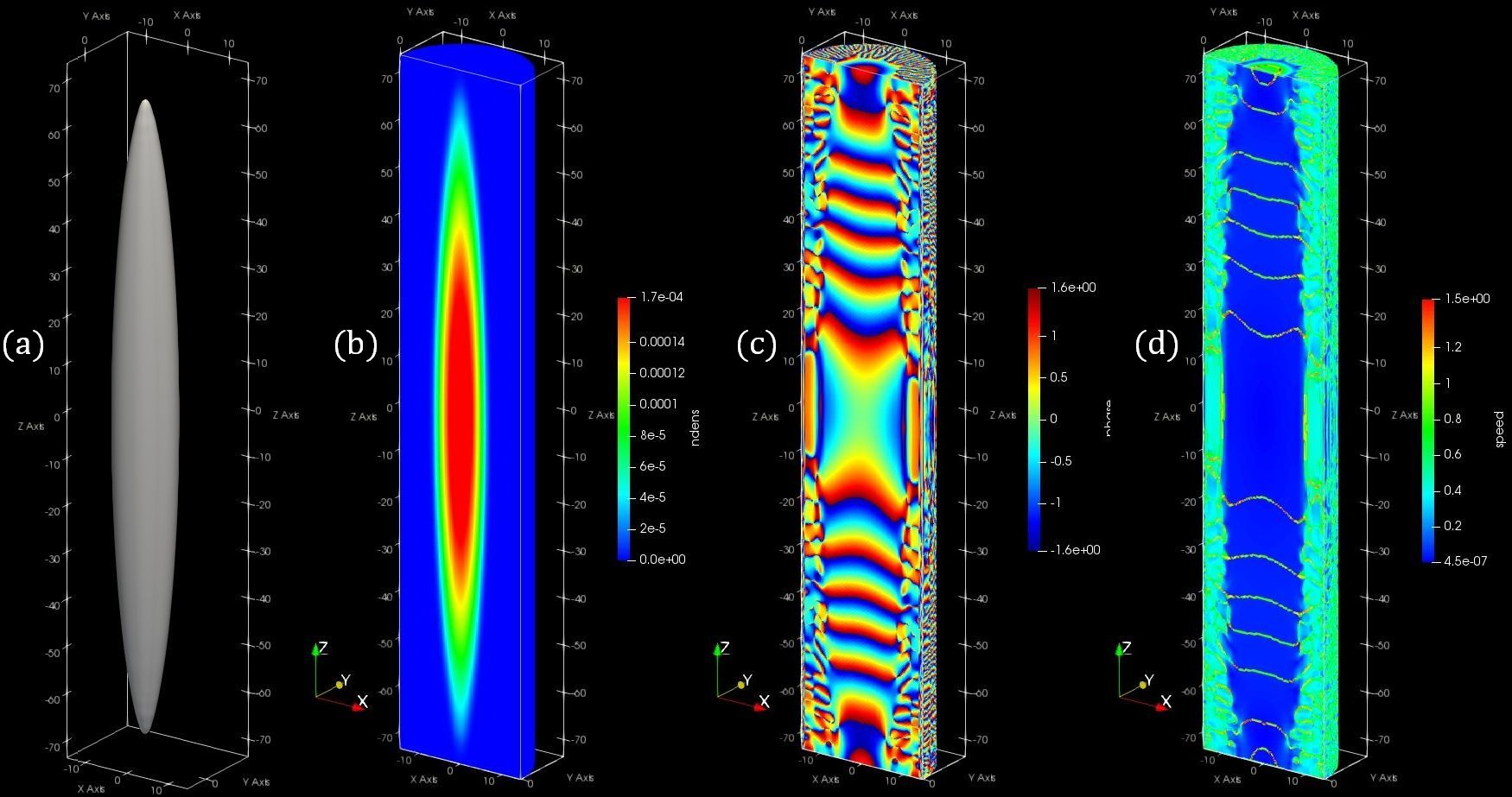}
\caption{Snapshot of the 3D contour of the condensate at $|\psi|^2=2.0\times 10^{-5}$ in (a) and its corresponding density, phase, and speed distribution on the cross-section at y = 0 and $t$ = 22 ms in (b), (c), and (d), respectively.}
\label{fig:Figure4}
\end{figure*}

\begin{figure*}[t]
\vspace{+0.5cm}
 \includegraphics[width=1.0\textwidth, clip, bb= 0 0 1750 923]{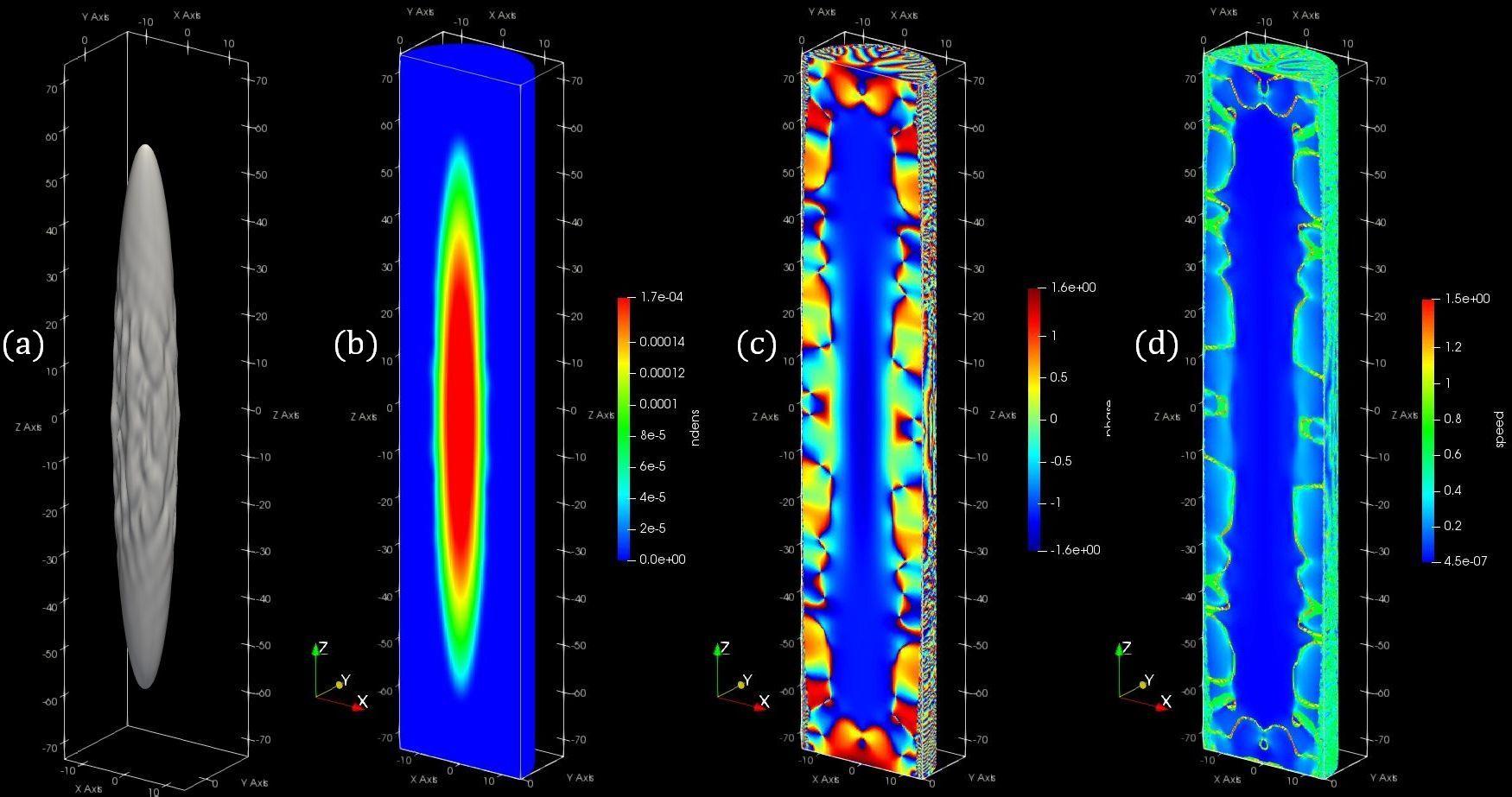}
\caption{Snapshot of the 3D contour of the condensate at $|\psi|^2=2.0\times 10^{-5}$ in (a) and its corresponding density, phase, and speed distribution on the cross section at y = 0 and $t$ = 91 ms in (b), (c), and (d), respectively.}
\label{fig:Figure5}
\end{figure*}

\begin{figure*}[t]
\vspace{+0.5cm}
 \includegraphics[width=1.0\textwidth, clip, bb= 0 0 1750 923]{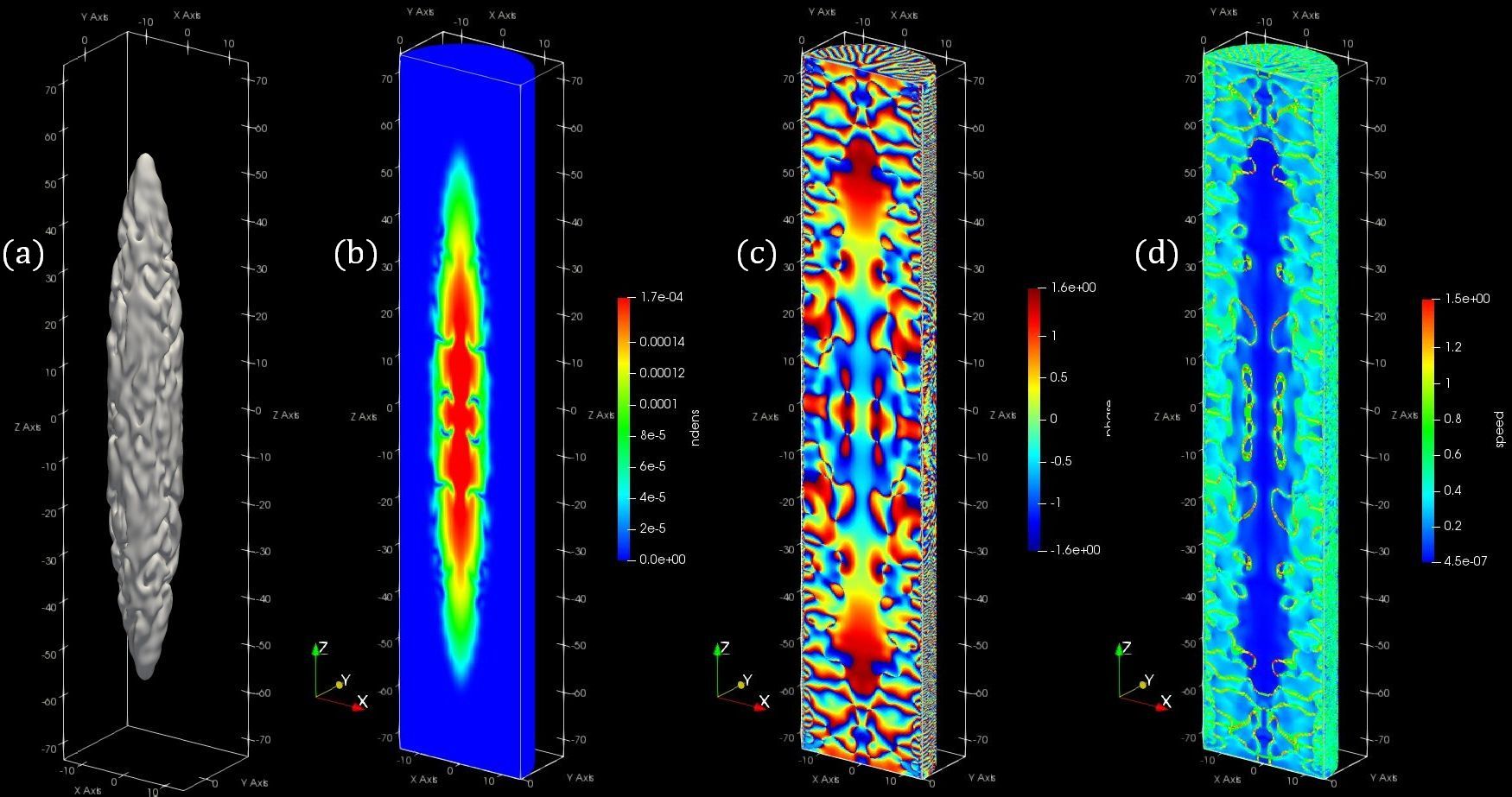}
\caption{Snapshot of the 3D contour of the condensate at $|\psi|^2=2.0\times 10^{-5}$ in (a) and its corresponding density, phase, and speed distribution on the cross-section at y = 0 and $t$ = 129 ms in (b), (c), and (d), respectively.}
\label{fig:Figure6}
\end{figure*}

\begin{figure*}[t]
\vspace{+0.5cm}
 \includegraphics[width=1.0\textwidth, clip, bb= 0 0 1750 923]{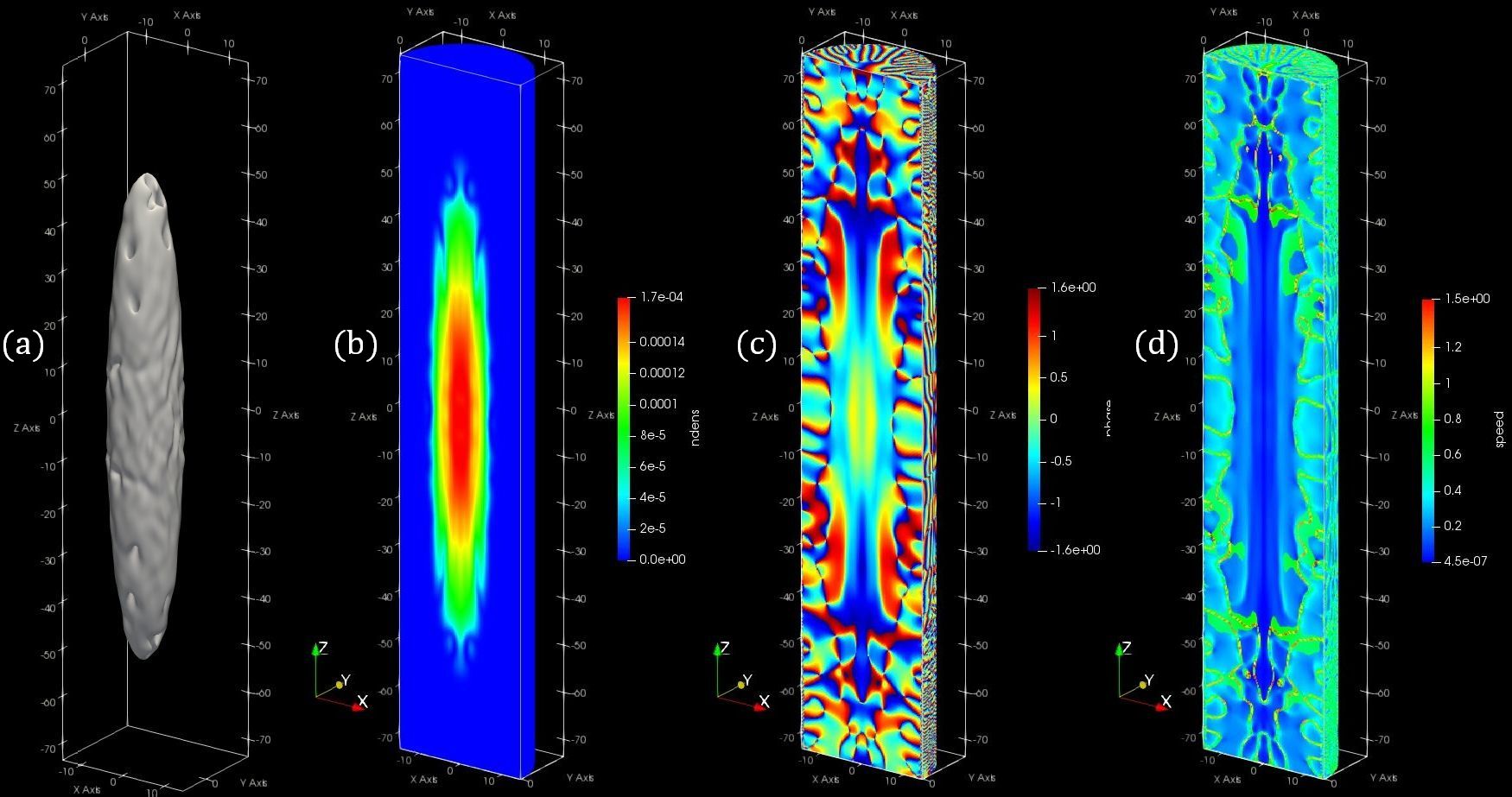}
\caption{Snapshot of the 3D contour of the condensate at $|\psi|^2=2.0\times 10^{-5}$ in (a) and its corresponding density, phase, and speed distribution on the cross section at y = 0 and $t$ = 195 ms in (b), (c), and (d), respectively.}
\label{fig:Figure7}
\end{figure*}

\begin{figure*}[t]
\vspace{+0.5cm}
 \includegraphics[width=1.0\textwidth, clip, bb= 0 0 1750 923]{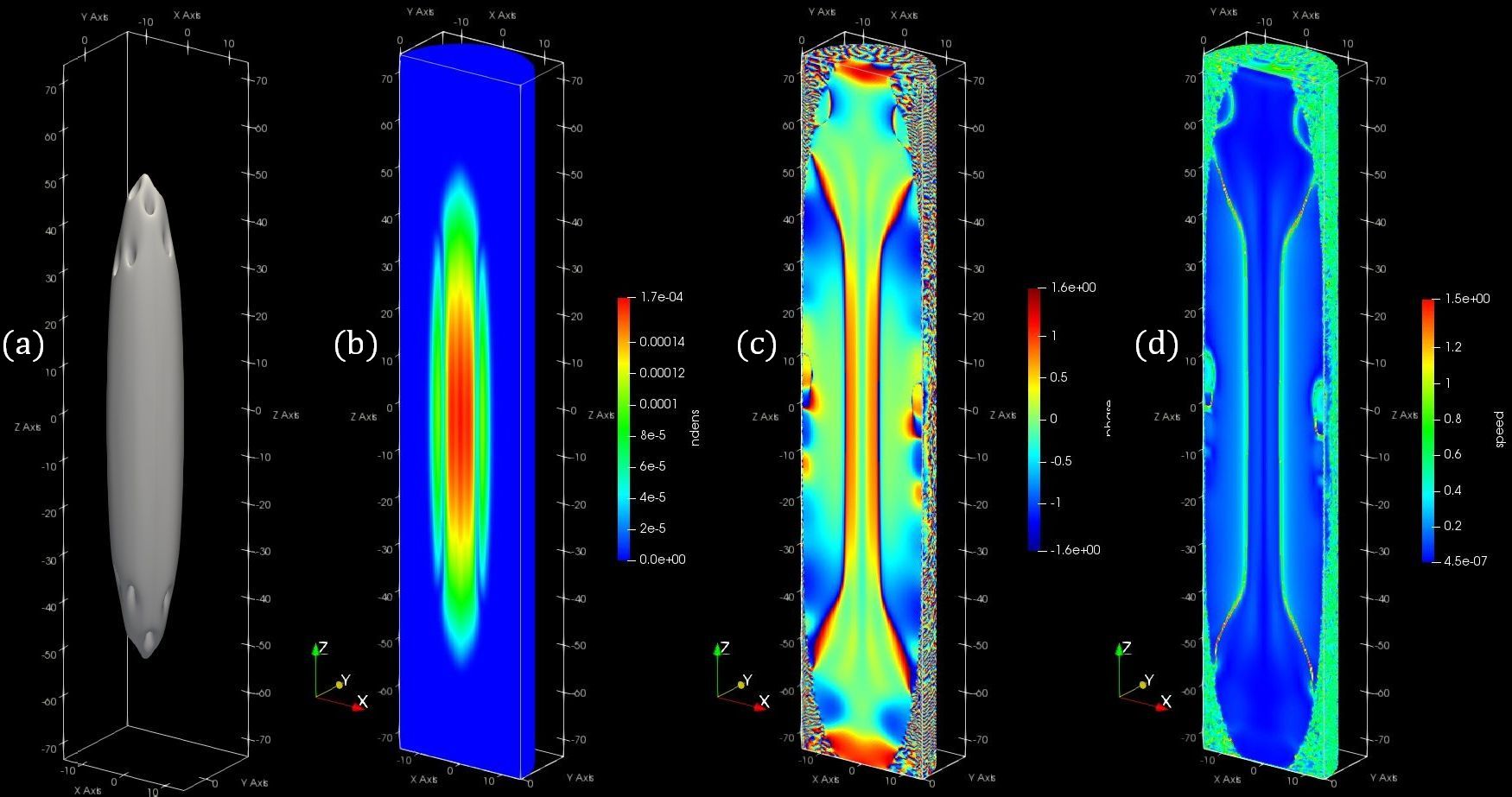}
\caption{Snapshot of the 3D contour of the condensate at $|\psi|^2=2.0\times 10^{-5}$ in (a) and its corresponding density, phase, and speed distribution on the cross-section at y = 0 and $t$ = 367 ms in (b), (c), and (d), respectively.}
\label{fig:Figure8}
\end{figure*}

Figures~\ref{fig:Figure4}-\ref{fig:Figure8} display a snapshot of the 3D contour of the condensate at $|\psi|^2=2.0\times 10^{-5}$ in (a) and its corresponding density, phase, and speed distribution on the cross-section at $\rm y = 0$ in (b), (c), and (d), respectively. The physical times shown in Figs.~\ref{fig:Figure4}-\ref{fig:Figure8} correspond to (a)-(d) of Fig.~\ref{fig:Figure1}. By comparing the contour plots in Figs.~\ref{fig:Figure4}-\ref{fig:Figure8}, we can confirm another key characteristic of the rotating BECs: the surface starts to fluctuate at approximately z = 0, forming vertically oriented stripes, which then propagate in the upper and lower vertical directions, causing turbulent density fluctuations on the surface. Subsequently, some of the bumps caused by the fluctuations grew into vortex lines that entered the condensate, whereas the remaining bumps disappeared because of dissipation. These results are consistent with those reported in Refs.~\cite{PhysRevA.71.063616}. A video of Figs.~\ref{fig:Figure4}-\ref{fig:Figure8} is available as supplementary material, which is provided as an auxiliary file for the preprint version.

In addition, Figs.~\ref{fig:Figure8}(b)-(d) further clarify the relationships between the density, phase, and speed distributions, as well as the spatial structure of the vortex lines. In particular, the position of the vertical hollow shown in Fig.~\ref{fig:Figure8}(b), that is, of a vortex line, corresponds to the edge of the vertical branch cut, as shown in Figs.~\ref{fig:Figure8}(c) and (d). Because the magnitude of the velocity was obtained as the absolute value of the phase gradient, the highlighted part of the speed distribution lies along the vortex. From Fig.~\ref{fig:Figure8}(c), we can observe the bending of the two vortex lines when they approach the inner edges of the condensate. This observation is consistent with those reported in Refs.~\cite{PhysRevA.71.063616, PhysRevLett.89.200403}. \HL{Note} that the \HL{resulting physical} time to form the lattice was approximately 1.5 to 2.0 times \HL{earlier} in our simulations than in the FD calculations in Ref.~\cite{PhysRevA.71.063616}. 
\HLJPCO{This may lead to a lack of reproduction of more detailed physics, e.g., the formation of scissors modes~\cite{PhysRevLett.83.4452} at the beginning of rotation, because the surface ripples appearing before the condensates have sufficiently experienced scissors oscillations.}
In SPH, the gradient of $\psi$ is given by the sum of $\psi$ in the neighboring areas weighted by the gradient of the kernel function. Therefore, the \HLJPCO{choice} of the kernel function affects the calculation accuracy. 
Further investigation is necessary to determine the optimal kernel function and required number of analytical particles.
\HLJPCO{In addition, SPH requires a reasonable number of particles to reproduce periodic oscillation phenomena, such as wave propagation, sufficiently accurately to avoid deviations in each period, as discussed in Section 5.2 in~\cite{Tsuzuki_2021}. Therefore, high-resolution computations are necessary to capture the physics in more detail. Consequently, we must implement our code on a multi-GPU platform, which will be performed in our future work.}
Nevertheless, our \HLJPCO{current} scheme can reproduce the key characteristic phenomenon of rotating 3D BECs in a 3D harmonic potential, forming a cigar-shaped condensate with a certain \HLJPCO{degree} of accuracy. 

\begin{figure*}[t]
\vspace{-39.0cm}
\includegraphics[width=4.55\textwidth, clip, bb= 0 0 3995 2486]{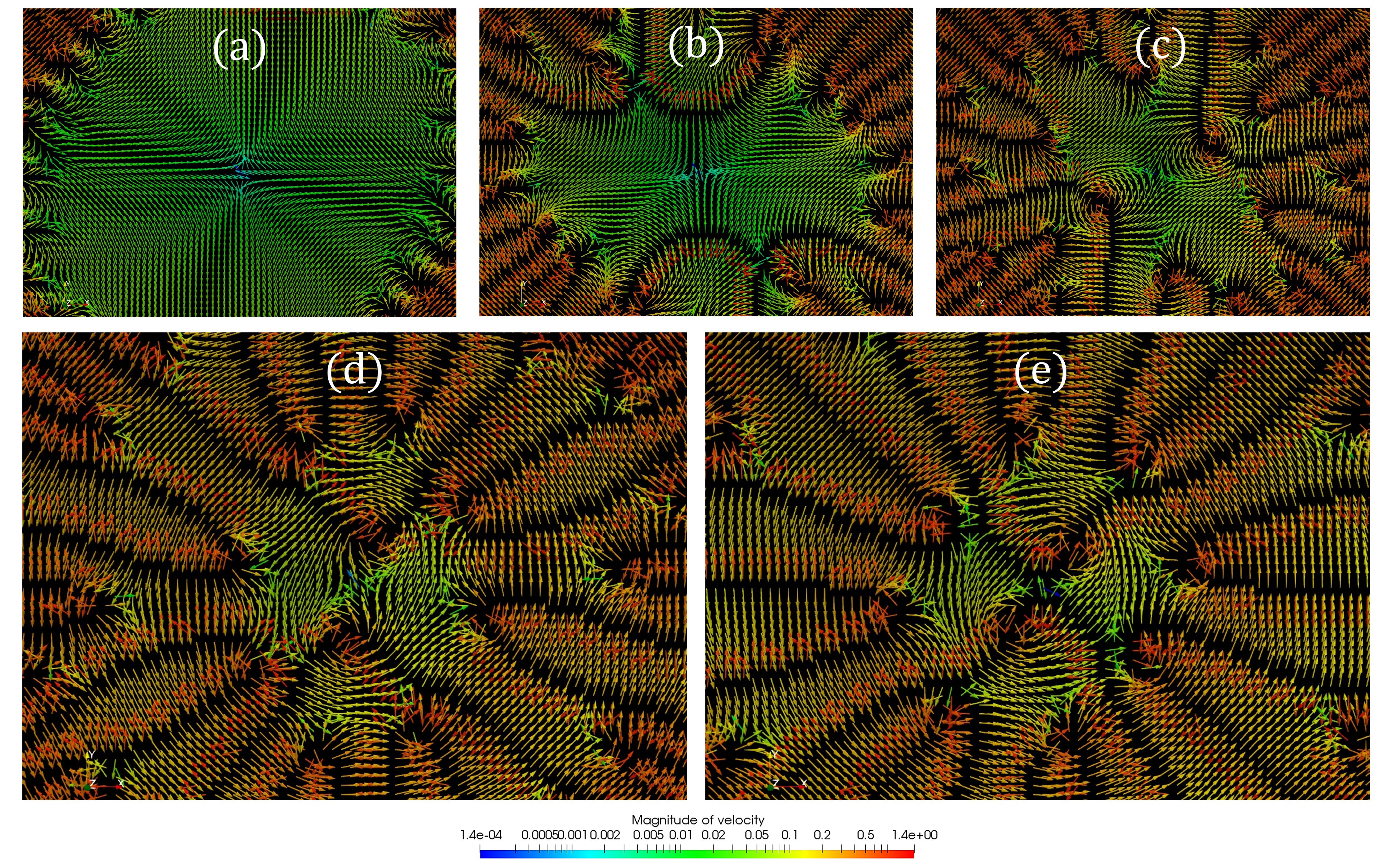}
\caption{The velocity vectors of the condensates on the cross-section at z = 0. These are obtained as the gradients of the phases at approximately (a) 22, (b) 91, (c) 129, (d) 195, and (e) 367 ms, corresponding to Fig.~\ref{fig:Figure1}(a)–(e) in the physical time.}

\label{fig:Figure9}
\end{figure*}

As mentioned above, we used a single GPU, and the total number of analytical particles was 9041088. We have confirmed that this number of particles per GPU can handle the entire physical process in a reasonable computation time. It follows that by increasing the number of GPUs in proportion to the number of particles, the total computation time can be kept constant as the resolution increases. At higher resolutions, our SPH scheme may help find the analytical solution of a point-vortex model for the spatial geometry of vortex lattices in 3D cases by exploring the behavior of ghost vortex lines in the vertical direction, similar to the 2D cases~\cite{PhysRevResearch.5.023109}. It may also be possible to compare the vortex lattice of the GP equation with that reproduced by a two-fluid model~\cite{doi:10.1063/5.0060605} to find a connection in the SPH form. 
\HLJPCO{In addition, Fig.~\ref{fig:Figure9} shows the velocity vectors of the condensates on the cross-section at z = 0. These are obtained as the gradients of the phases at approximately (a) 22, (b) 91, (c) 129, (d) 195, and (e) 367 ms, corresponding to Fig.~\ref{fig:Figure1}(a)-(e) in the physical time. A video of Fig.~\ref{fig:Figure9} is provided as a supplementary material, which is available as an auxiliary file with the preprint version. The lengths of the vectors are kept the same for visualization; however, they are colored according to the magnitude of the velocity on the logarithmic scale. We observe that the external vortices in the outer region flow along the velocity fields in the latter part of the simulation after the vortex grid is formed. Figure 9 shows the kinematic and mechanical images of the vortices corresponding to the phase singularities in mathematics. The stripes correspond to the branches cut in the phase where the velocity fields are undefined. Further fine-grained high-resolution calculations may capture the internal dynamics inside the vortices, such as the helical structure of the Kelvin waves~\cite{thomson_1880}.} For these reasons, a multi-GPU implementation would be necessary in future studies.

\section{Conclusion}
Recently, we presented a new numerical scheme for vortex lattice formation in a rotating BEC using SPH with an explicit time-integrating scheme; our SPH scheme quantitatively reproduced vortex lattices and their formation processes in quasi-2D rotating BECs trapped in a 2D harmonic potential. 
This study successfully demonstrated a 3D simulation of rotating BECs trapped in the 3D harmonic potential, forming cigar-shaped condensates. We found that our scheme can reproduce the following typical behaviors of rotating 3D BECs reported in the literature: (i) the characteristic
shape of the lattice formed in the cross-section at the origin and its formation process, (ii) the stable existence of vortex lines in the vertical direction after reaching the steady state, (iii) a ``cookie-cutter'' shape, where a similar vortex lattice is observed wherever the condensate is cut horizontally in a certain vertical region, (iv) the bending of the vortex line when approaching the inner edges of the condensate, and (v) the formation of vortex lattices by vortices entering from outside the condensate. We further demonstrated our scheme by simulating rotating 3D BECs.

SPH has significant potential for analyzing problems in low-temperature physics, such as quantum fluids such as liquid helium-4 and ultracold dilute gases, because the mechanisms of SPH and real BEC physics provide a mechanical picture of many-particle interactions. However, limited research has been conducted on the application of SPH to address these problems. We hope that this study will help make SPH widely accepted in low-temperature physics and will be used to address these problems.

\section*{Acknowledgment}
This study was supported by JSPS KAKENHI Grant Number 22K14177. 
The authors thank Editage (www.editage.jp) for the English language editing.

\section*{Data availability statement}
No new data were created or analysed in this study.

\bibliographystyle{h-physrev3}
\bibliography{reference}

\end{document}